\documentclass{article}
\PassOptionsToPackage{numbers,sort&compress}{natbib}
\usepackage[main, final]{neurips_2026}
\usepackage{subcaption}
\usepackage{amsmath}
\usepackage{comment}
\usepackage{xspace}
\usepackage[table]{xcolor}
\usepackage{appendix}
\usepackage{amsmath}
\usepackage{amssymb}  % Extra maths symbols
\usepackage{physics}
\usepackage{comment}
\setcitestyle{square}

\let\<\langle
\let\>\rangle

\DeclareMathOperator{\sign}{sign}

\newcommand{\xfg}{$X_{f,\text{gen}}$}
\newcommand{\xfd}{$X_{f,\text{data}}$}
\renewcommand{\xi}{$X_i$}
\newcommand{\ndensity}{$n^*$}
\newcommand{\inte}{$u^*$}

\newcommand{\err}{$\delta_{\text{err}}$}
\newcommand{\jet}{$L_{\text{jet}}$}

\usepackage[utf8]{inputenc} % allow utf-8 input
\usepackage[T1]{fontenc}    % use 8-bit T1 fonts
\usepackage{hyperref}       % hyperlinks
\usepackage{url}            % simple URL typesetting
\usepackage{booktabs}       % professional-quality tables
\usepackage{amsfonts}       % blackboard math symbols
\usepackage{nicefrac}       % compact symbols for 1/2, etc.
\usepackage{microtype}      % microtypography
\usepackage{xcolor}         % colors
\usepackage{graphicx}

\title{Learning to Jet: A Surrogate Subgrid Model for AGN Feedback in Cosmological Simulations}
\workshoptitle{Machine Learning and the Physical Sciences}

\author{%
  Leena Iwamoto \\
  Harvard University \\
  Cambridge, MA 02138 \\
  \texttt{liwamoto@college.harvard.edu}
  \And
  Olga Borodina \\
  Center for Astrophysics | Harvard \& Smithsonian \\
  Cambridge, MA 02138 \\
  \texttt{olga.borodina@cfa.harvard.edu}
  \And
  Mark Goldstein \\
  Center for Computational Mathematics \\
  Flatiron Institute \\
  New York, NY 10010 \\
  \texttt{mgoldstein@flatironinstitute.org}
  \And
  Carolina Cuesta-Lazaro \\
  Center for Computational Astrophysics\\
  Flatiron Institute \\
  New York, NY 10010 \\
  \texttt{ccuesta-lazaro@flatironinstitute.org}\\
  \And
  Keiya Hirashima \\
Center for Interdisciplinary Theoretical \\
and Mathematical Sciences RIKEN \\
Wako, Japan\\
  \texttt{keiya.hirashima@riken.jp}\\
  \And
  Lars Hernquist \\
  Center for Astrophysics | Harvard \& Smithsonian \\
  Cambridge, MA 02138 \\
  \texttt{lhernquist@cfa.harvard.edu}
}

\begin{document}

\maketitle

\begin{abstract}
Cosmological simulations demonstrate the importance of active galactic nuclei (AGN) in quenching star formation and shaping the long-term evolution of galaxies. However, their limited resolution prevents them from directly capturing many important physical processes, including interactions between AGN jets and the surrounding interstellar medium (ISM). Here, we introduce a novel method for modeling AGN jet feedback: a generative flow-matching model with a 3D U-Net architecture, trained to predict AGN jet evolution. Our model is trained on pairs of coarse-grained, high-resolution AGN jet simulations that resolve the turbulent ISM within the central 1\,kpc of the host galaxy. The model reproduces the spatial distribution of five key physical quantities and recovers the total energy with a median error of 13.5\% over held-out validation simulations. During model rollout, or repeated application of the model to its own outputs, it remains stable over several steps, with final total-energy errors below 10\% in the best-performing regimes. At around three orders of magnitude faster than direct simulation, our model functions as a viable subgrid prescription for AGN jet feedback in cosmological simulations.
\end{abstract}

\section{Introduction}

\paragraph{Background.} Active Galactic Nuclei (AGN) play a critical role in quenching star formation in galaxies and driving their evolution over cosmic time \citep{Di_Matteo_2005, croton2006, Fabian2012}. AGN jets are powerful streams of electrons and baryons launched from the center of galaxies \citep{blandford}, depositing energy into the surrounding interstellar (ISM), circumgalactic, and intracluster medium \cite{coevolution2013}. Jet propagation occurs on a range of scales, from subparsec to megaparsec distances relative to the central Supermassive Black Hole \citep[SMBH;][]{agnreview2020}. Given that AGN jet feedback is a process that spans many orders of magnitude in physical scale, it is challenging to capture in a single simulation.

\paragraph{Challenge.} Large-scale hydrodynamical simulations have emerged as important tools to simulate the cosmological history of the universe \cite{tng, eagle, simba}. Simulations run from identical initial conditions but with different AGN feedback prescriptions produce widely different galaxy and gas properties, underscoring the impact of the chosen feedback model \citep{camels}. Because resolving all AGN propagation scales is computationally infeasible, state-of-the-art simulations such as IllustrisTNG \citep[TNG;][]{tng}, EAGLE \citep{eagle}, and SIMBA \citep{simba} necessarily replace the jet-mode with their own subgrid prescriptions. However, none of these simulations resolve the coupling between the AGN jet and turbulent ISM: jets can be slowed down, redirected, or even stalled by the dense medium \citep{olga}. As such, unresolved jet-ISM interactions in cosmological simulations can substantially alter jet propagation, and further, larger-scale galaxy evolution.

\paragraph{Surrogate models.} Neural Networks (NNs) provide a compelling alternative to the current, heuristic prescriptions for unresolved AGN feedback in cosmological simulations. Rather than approximating small-scale physics with an ad-hoc solution, a surrogate model can be trained on the behavior of expensive, high-resolution simulations to emulate unresolved physics at a fraction of the computational cost. Recent work has developed a NN-based surrogate model for supernova feedback \cite{hirashima}, and a neural-operator surrogate for black hole accretion-driven feedback trained on high-resolution general-relativistic simulations \cite{grmhdsurrogate}. Building on this framework, our project extends the surrogate modeling approach to AGN jets, aiming to learn their evolution for integration into cosmological simulations.

\paragraph{Contribution.} In our work, we train a conditional flow-matching model with a 3D U-Net architecture on detailed AGN simulations \cite{olga} to predict the evolution of AGN jets. In learning to reproduce jet evolution, the surrogate model avoids the computational cost of resolving jets in the simulation itself. By training on realistic jet simulations, the model also provides a more physically grounded alternative to the feedback prescriptions used in current cosmological simulations. We ultimately aim to incorporate our validated surrogate into the next generation of such simulations, enabling a more physically motivated treatment of AGN jet feedback.

\begin{figure}[t]
  \centering
  \begin{subfigure}[b]{0.23\linewidth}
    \includegraphics[width=\linewidth]{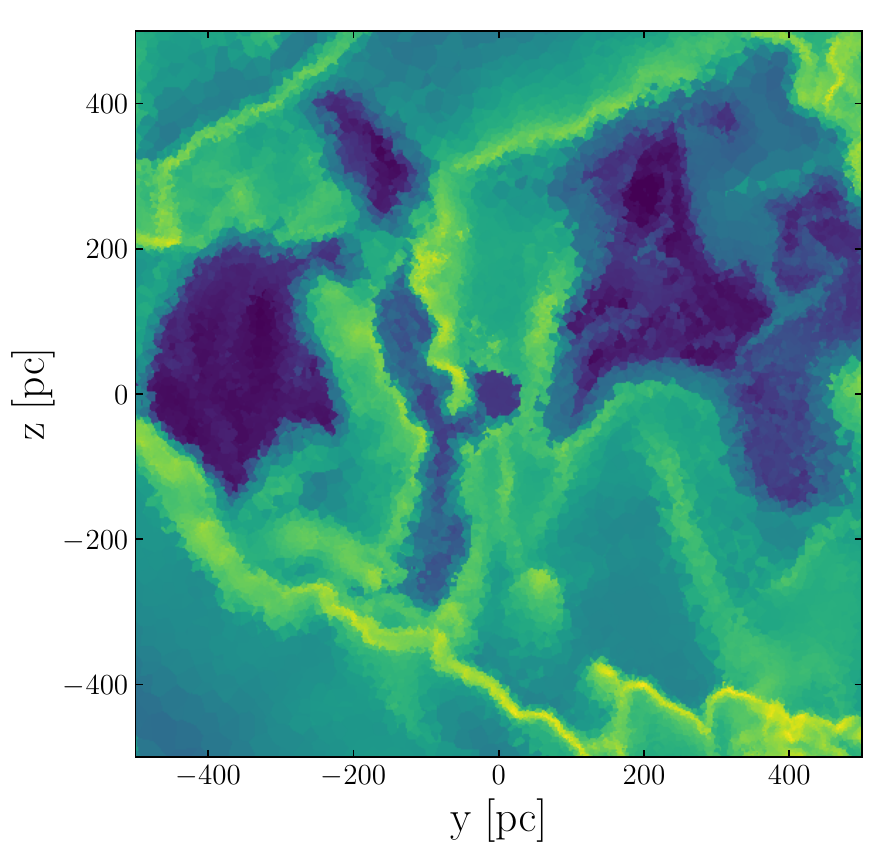}
    \caption{AREPO, native}\label{fig:jet-native}
  \end{subfigure}\hfill
  \begin{subfigure}[b]{0.23\linewidth}
    \includegraphics[width=\linewidth]{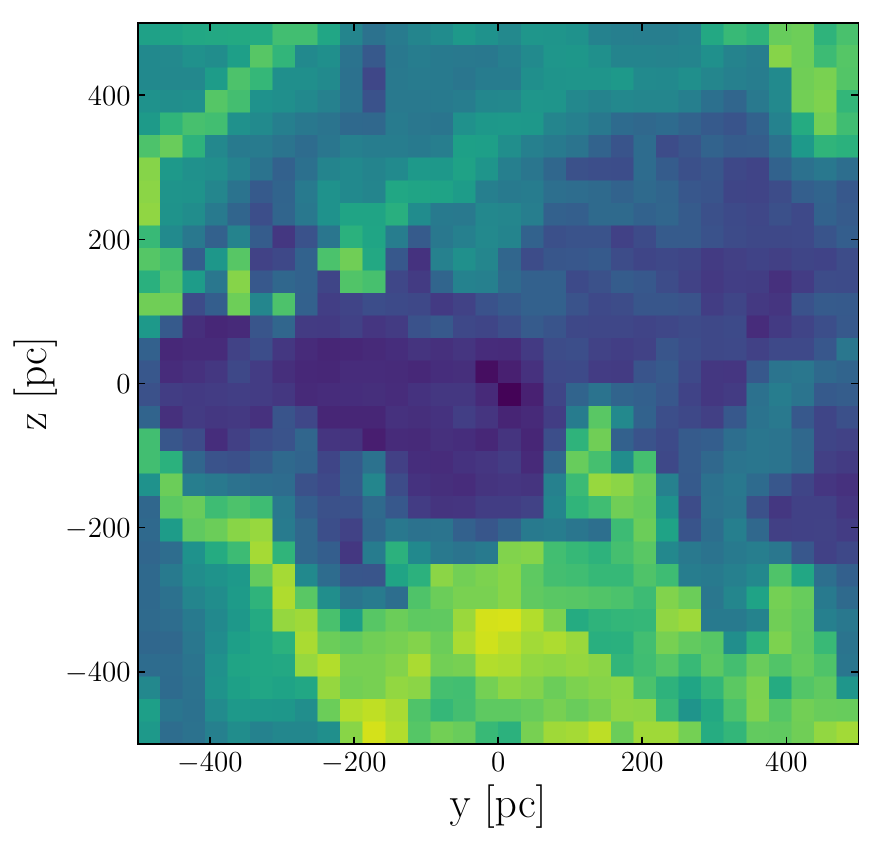}
    \caption{$32$ cells/kpc}\label{fig:jet-32}
  \end{subfigure}\hfill
  \begin{subfigure}[b]{0.23\linewidth}
    \includegraphics[width=\linewidth]{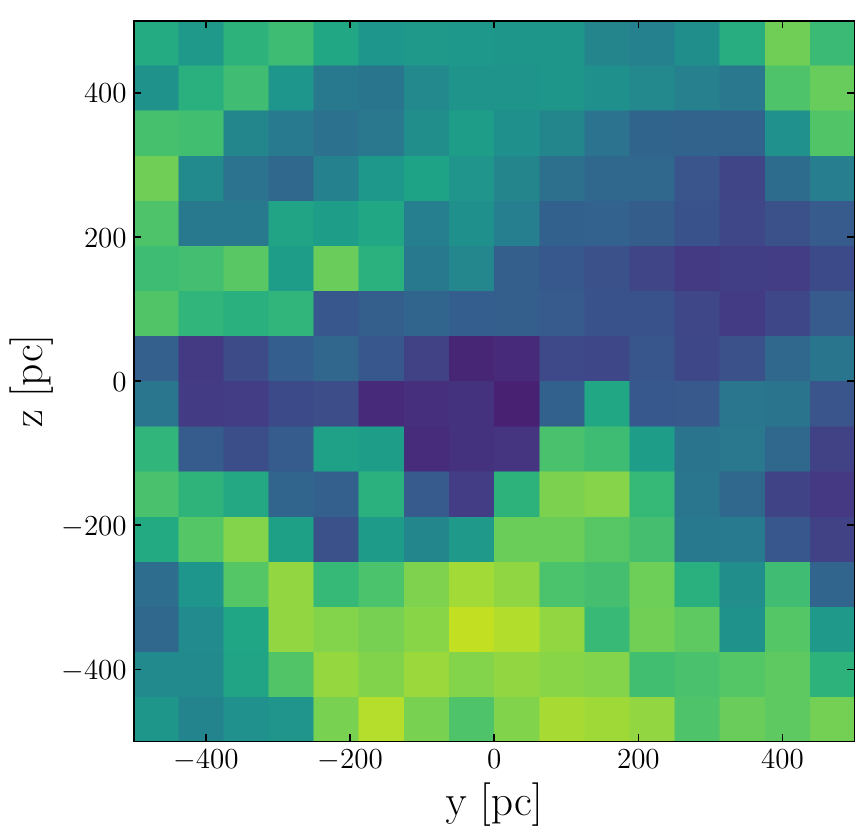}
    \caption{$16$ cells/kpc}\label{fig:jet-16}
  \end{subfigure}\hfill
  \begin{subfigure}[b]{0.3\linewidth}
    \includegraphics[width=\linewidth]{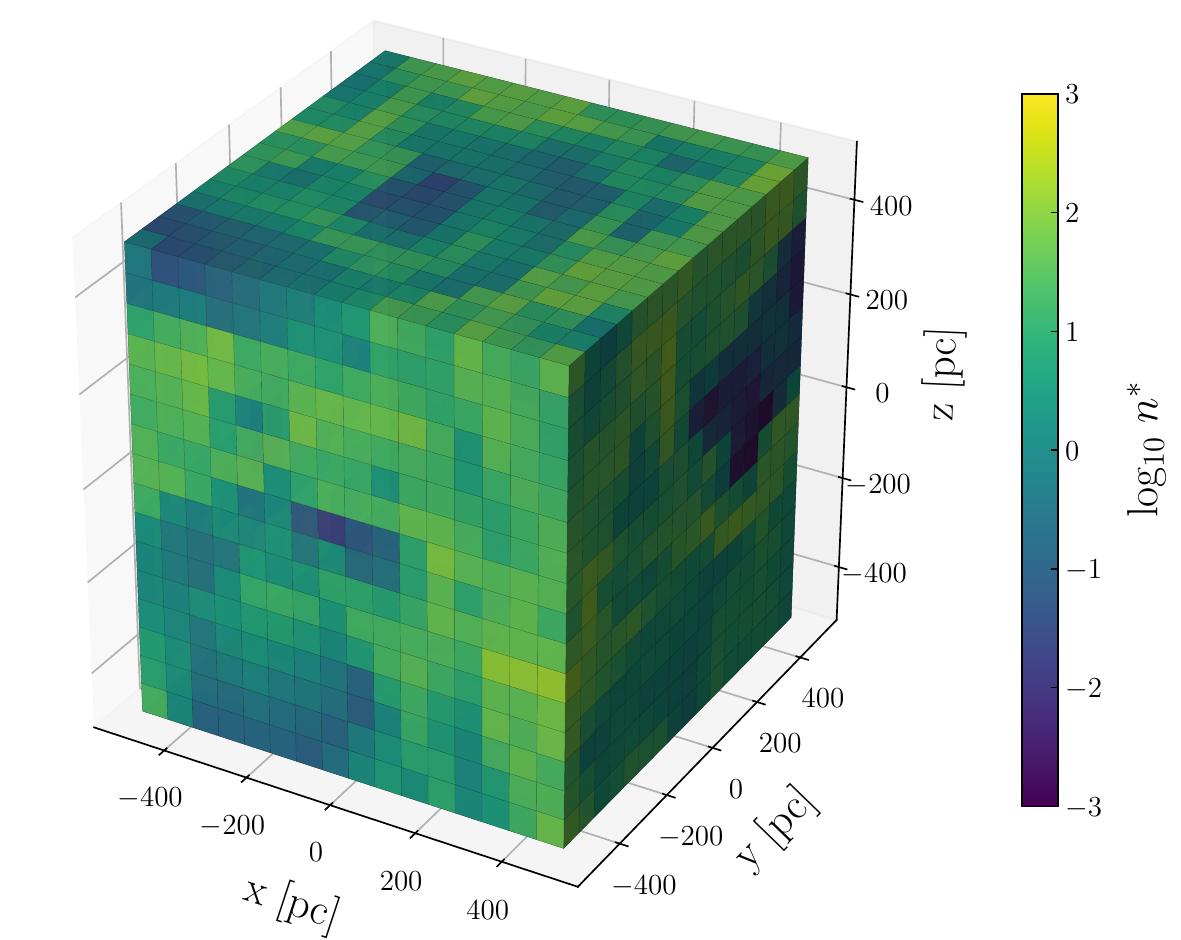}
    \caption{$16^3$, 3D view}\label{fig:jet-3d}
  \end{subfigure}
  \caption{Coarse-graining the data. (a) Number density from the original AREPO simulations \citep{olga} with box side length 1\,kpc.
  (b, c) The same snapshot resampled from the Voronoi mesh onto a regular grid with $32^3$ and $16^3$ cells, respectively. We train at $16^3$, as $1/16$\,kpc per cell  matches the best resolution near black holes in cosmological simulations. (d) 3-D rendering of the $16^3$ field.}
  \label{fig:preprocessjet}
\end{figure}

\section{Method}
\paragraph{Training data.} Detailed simulations of AGN jets and the surrounding turbulent ISM \cite{olga} serve as the training dataset for our machine learning model. They were simulated using the Arepo code \citep{arepo}, solving hydrodynamic equations on a moving Voronoi mesh. Our dataset contains 3659 snapshots in timesteps of $\sim$0.05\,Myr, yielding $\sim$100 thousand pairs of snapshots that cover a wide variety of physical scenarios. Each simulation scenario is defined by a set of initial conditions, mainly the number density and Mach number of the turbulent medium, ranging between [0.1, 100]\,cm$^{-3}$ and [0, 6], respectively. The jet power is injected at a constant rate along a fixed axis, with power ranging between $10^{37.4}$--$ 10^{44.5}$\,erg\,s$^{-1}$. We limit training to the central (1\,kpc)$^3$ region because it contains the most physically significant zone of AGN jet-ISM interactions. The training and validation sets contain 50 and 7 unique simulations, respectively. We partition at the simulation level before pairing snapshots such that every snapshot falls in a single split, preventing any overlap between the training and validation data. For more details on the dataset, including the array of training and validation simulation initial conditions, see Appendix \ref{sec:appendixb}.

\paragraph{Data preprocessing.} 
Due to the irregular, Voronoi mesh structure of the raw simulation data, the first step in our preprocessing algorithm was to transform the data to a 3-D Cartesian grid structure. Thus, we took the 1\,kpc central region and split it up into a $16\times16\times16$ gridded cube. We chose 16 as our coarse-graining size as it approximately matches the best resolution of star-forming galaxies in TNG, coarse-graining to ease the implementation in future cosmological simulations. This coarse-graining inevitably removes fine-scale structure, as illustrated in Fig. \ref{fig:preprocessjet}.

To obtain the physical quantity values for each cell in our 3-D grid, we developed an algorithm that identifies all Voronoi mesh-generating particles residing within a given grid-cell and averages their corresponding quantities. In total, we extracted five physical quantities per grid-cell: number density $n$, specific internal energy $u$, and 3 velocity components $v_j$; for all equations, see Appx. \ref{sec:appendixa}. It was critical that the coarse-graining process conserved mass, momentum, and total energy, which we ensured by using volume and mass-weighted average, respectively, with a subsampling correction for cells containing no particle (Appx. \ref{sec:appendixa}). Our five final physical quantities are thus summarized as follows:
\begin{align}\label{eq:params}
    P \in \{n^*, u^*, v^*_{x}, v^*_{y}, v^*_{z}\},
\end{align}
where $n^*$ is normalized number density, $u^*$ is the normalized internal energy, and $v^*_{j}$ are the Cartesian velocities, with $j\in[x,y,z]$.

\begin{figure}[t]
    \centering
    \includegraphics[width=\linewidth]{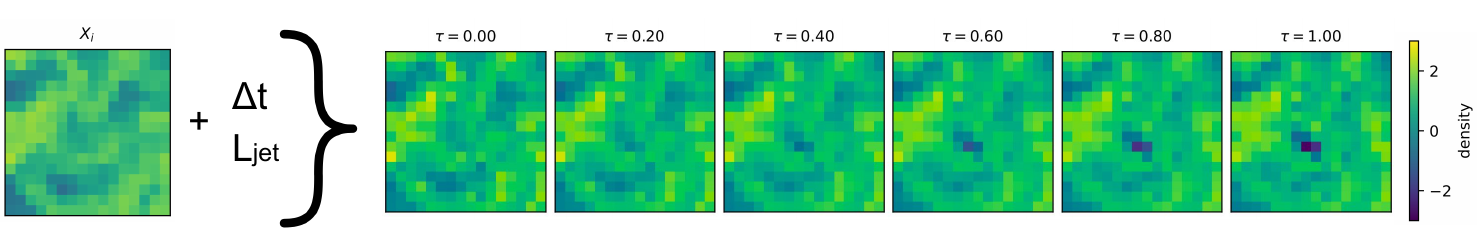}
    \caption{The conditional flow-matching task. Conditioned on the clean initial snapshot $X_i$, the time gap $\Delta t$, and the jet power \jet\ (left of bracket), the network $v_\theta$ transports the noised state $z_0 = X_i + \lambda\mathcal{N}(0,1)$ to the final snapshot $z_1 = X_f$. Panels show the interpolant $z_\tau$ at six flow-matching times $\tau$, on a central slice of the normalized density channel.}
    \label{fig:flowmatching}
\end{figure}

\paragraph{Computational problem.} We use flow matching to sample the \textit{final snapshot}, $X_f$, given the \textit{initial snapshot}, $X_i$. This is done by learning a velocity field $v_\theta$, conditioned on the clean initial snapshot $X_i$, that transports samples from a base distribution to the target distribution, $z_1 = X_f$. Following \cite{chen2024probabilistic}, the base sample is a noised initial snapshot, $z_0 = X_i + \lambda\varepsilon$, with $\varepsilon \sim \mathcal{N}(0, I)$ and $\lambda = 0.062$. Integrating $v_\theta$ from $\tau = 0$ to $\tau = 1$ moves $z_0$ along a path $z_\tau$ to $z_1$. We optimize a flow-matching loss of the form
\begin{align}
    \mathcal{L}_{\text{FM}}  = \sum_{\text{channels}}  \big||v_\theta(z_\tau, \tau, \text{flags}) - (z_1-z_0)|\big|^2 
\end{align}
where the `flags' are the conditioning inputs: one tensor, the clean initial snapshot $X_i$, and two scalars, the physical time difference $\Delta t$ between snapshots and the constant launched jet power, \jet. The linear stochastic interpolant is of the classic flow-matching form $z_\tau = (1-\tau) z_0 + \tau z_1,$
where $\tau \in [0,1]$. A visualization of the computational problem is provided in Fig. \ref{fig:flowmatching}.

\paragraph{Architecture.} We employ a 3D U-Net architecture \citep{unet} to parameterize $v_\theta$. The network acts directly on volumetric data using isotropic 3D convolutions, allowing the model to capture spatial correlations between quantities, especially necessary to learn the correlations in jet propagation. The encoder and decoder are connected via U-Net skip connections, with a self-attention layer in the bottleneck to capture long-range spatial correlations. The conditioning flags are injected into every residual block via FiLM-style \citep[]{film} feature-wise scale-and-shift.

\section{Results}\label{sec:results}
\paragraph{Computational cost.} The high-resolution AREPO jet simulations in our dataset require a median of $\approx40$ CPU-hours per $\Delta t = 0.05$\,Myr timestep. Once trained, our model predicts the evolved state of the 1\,kpc region in $\approx0.05$ CPU-hours per $\Delta t$, a reduction of roughly three orders of magnitude. This is the regime required for a subgrid model to be called at every relevant timestep of a cosmological simulation. Further, unlike direct hydrodynamical simulations, the cost of a model evaluation is independent of the given time interval; thus, the model can advance the region across a full cosmological timestep in one $\approx0.05$ CPU-hour evaluation.

\begin{figure}
  \centering
  \begin{subfigure}[b]{0.52\linewidth}
    \includegraphics[width=\linewidth]{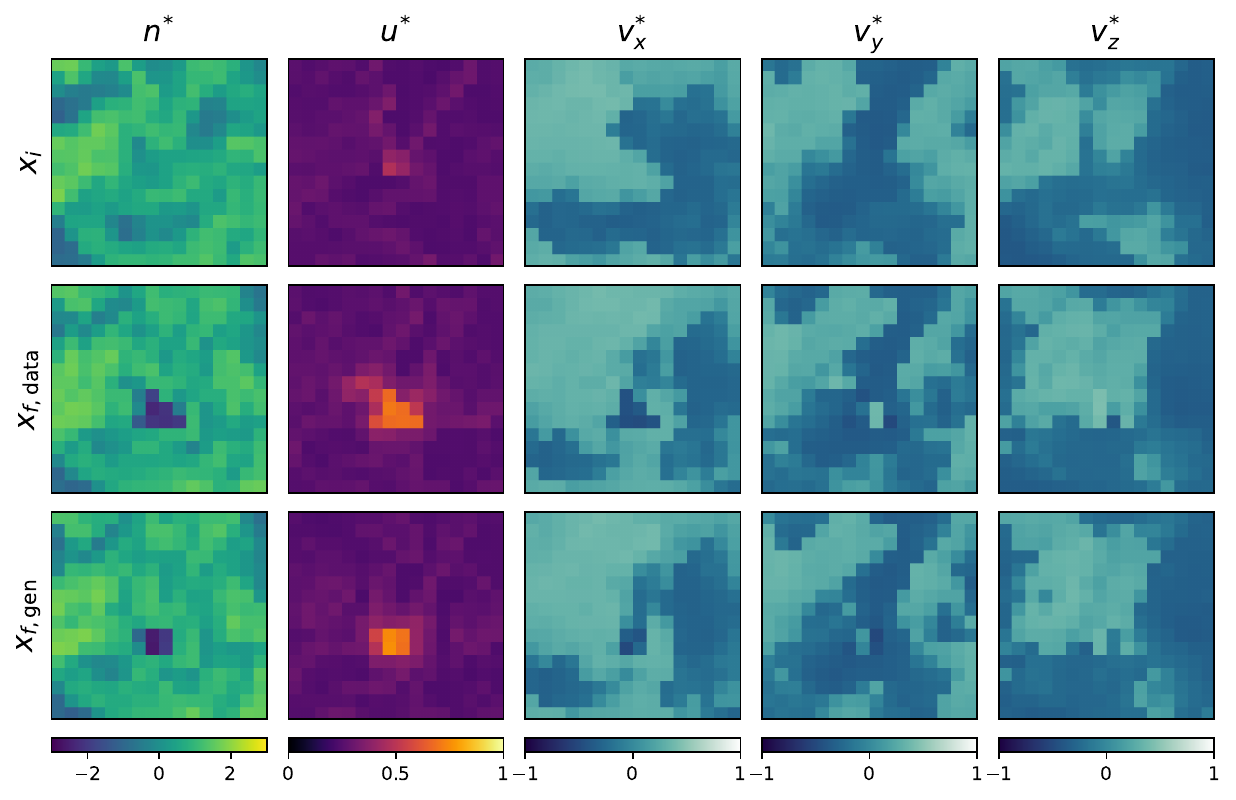}
    \caption{Physical quantities}\label{fig:valslices}
  \end{subfigure}\hfill
  \begin{subfigure}[b]{0.48\linewidth}
    \includegraphics[width=\linewidth]{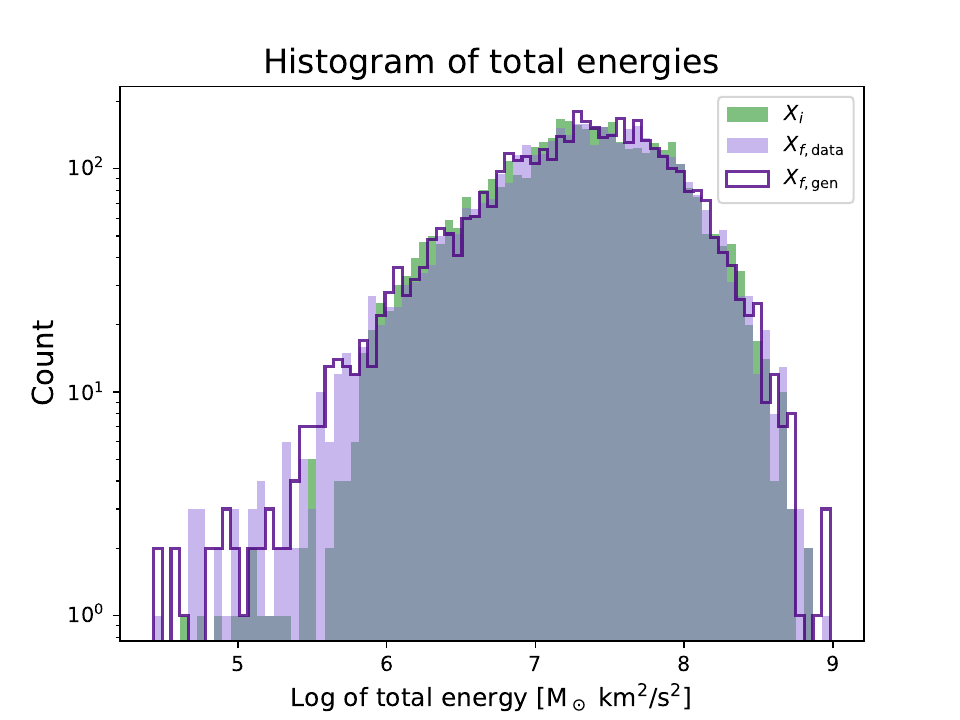}
    \caption{Energy distributions}\label{fig:energydists}
  \end{subfigure}
  \caption{Results for $n=10\,\text{cm}^{-3}, L_{\text{jet}} = 10^{40}\,\text{erg}/\text{s}, \mathcal{M} =2$ validation simulation. (a) Top row: Initial held-out validation snapshot \xi. Middle row: ground truth (\xfd) in $\Delta t = 1.5$\,Myr. Bottom row: model generated (\xfg). Columns correspond to the physical quantities of normalized number density \ndensity, internal energy \inte, and the three velocity components $v_j^*$, with a per-channel MSE of $\delta_{\text{err}} (\%) = [ 0.23, 0.26, 0.28, 0.31, 0.30]$. (b) Total energy distributions for the same validation pair as plotted in (a), with an energy percent error of $\varepsilon_{\text{err}}=2.70$\%.}
  \label{fig:params}
\end{figure}

\paragraph {Physical quantity reconstructions.}  We find that the trained model is able to predict the distribution of each physical quantity (Eq. \ref{eq:params}) for the final snapshot, $X_f$, with high fidelity, for held-out validation data. In Fig. \ref{fig:valslices}, we display slice plots of number density $n^*$, internal energy $u^*$, and the three velocity components $v_j^*$, respectively. We define a mean-squared-error metric,
\begin{align}\label{eq:deltaerr}
    \delta_{\text{err}} = \frac{1}{N}\sum_{i=1}^N \left(X_{\text{gen},i} - X_{\text{final},i}\right)^2,
\end{align}
where here, $i$ is a given cell, and $\delta_{\text{err}}$ is reported multiplied by 100 throughout. Over a single timestep of 1.5\,Myr between initial and final snapshots, the model is able to predict \xfg\ with high fidelity, yielding a model-space per-quantity $\delta_{\text{err}} (\%)= [ 0.23, 0.26, 0.28, 0.31, 0.30]$. The general morphology and intensity range of each quantity are represented well by our model's prediction. Averaged over all validation pairs (five model realizations each) and over the five physical quantities, $\delta_{\text{err}} = 0.94 \pm 1.0 \%$ (in model space), where the uncertainty is the standard deviation across pairs and realizations.

\paragraph{Energy reconstructions.} It is crucial to be able to predict the coupling of the jet energy to the ISM. Thus, we need to reconstruct not only the separate physical quantities, but also the overall energy distribution. Fig. \ref{fig:energydists} shows the total energy distributions of a given $X_i$ and \xfd\ snapshot pair compared to the model's prediction, \xfg. The overall alignment of the \xfd\ and \xfg\ energy distributions demonstrates the model's ability to reconstruct not only the bare physical quantities on which it was explicitly trained, but also the total energy distributions. This alignment is quantified by 
\begin{align}\label{eq:energyerr}
    \varepsilon_{\text{err}} = (E_{\text{gen}}-E_{f})/E_f, 
\end{align}
or the percent error between model and data total energies, where $E_{\text{gen}}$ is the model-generated energy for the final snapshot $X_f$, and $E_f$ is the real energy for $X_f$, yielding $\varepsilon_{\text{err}} = 2.7\%$. Averaged over all validation pairs (five model realizations each), the absolute total-energy error is $|\varepsilon_{\text{err}}| = 19.6 \pm 34.5\%$ (median $13.5\%$), which is heavily dominated by the errors in two simulations, with the rest of the validation simulations averaging $|\varepsilon_{\text{err}}| \leq 12\%$. For more details, see Appendix \ref{sec:onestep}

\begin{figure}
  \centering
    \includegraphics[width=\linewidth]{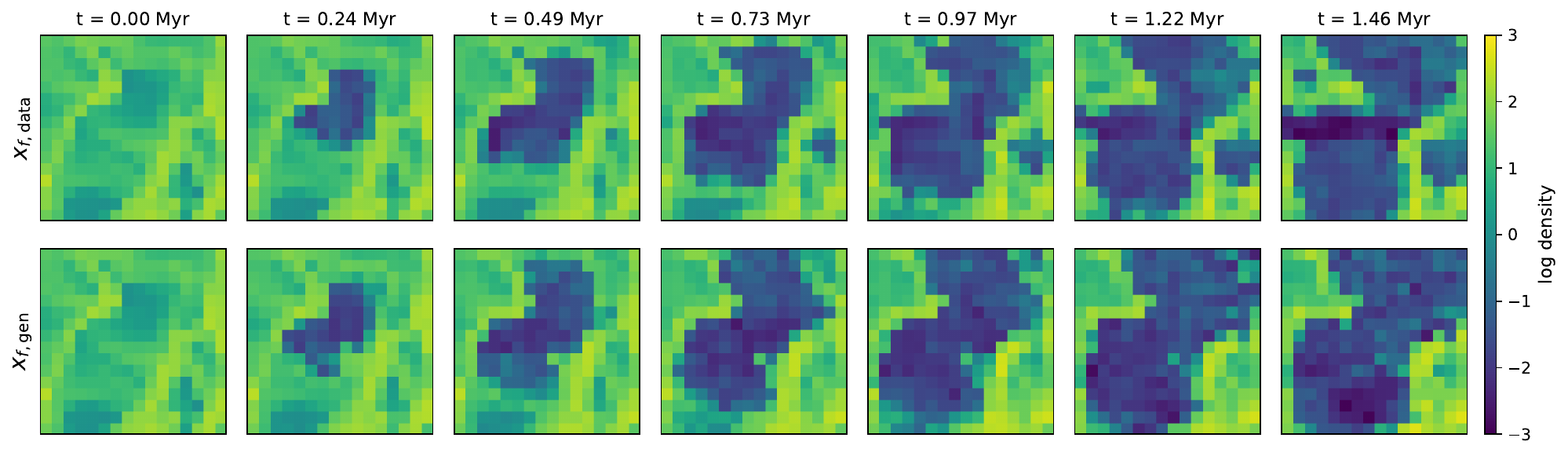}
  \caption{Rollout for $n=60\,\text{cm}^{-3}, L_{\text{jet}} = 10^{43}\,\text{erg}/\text{s}, \mathcal{M} =3$ validation simulation. Over 6 steps of $\Delta t \approx0.25$\,Myr, the model feeds its own previous output back in as \xi\ to predict the next snapshot, \xfg. For rollout of other physical quantities and validation simulations, see Appendix \ref{sec:appendixrollout}.}
  \label{fig:rollouts}
\end{figure}

\begin{figure}
  \centering
    \includegraphics[width=\linewidth]{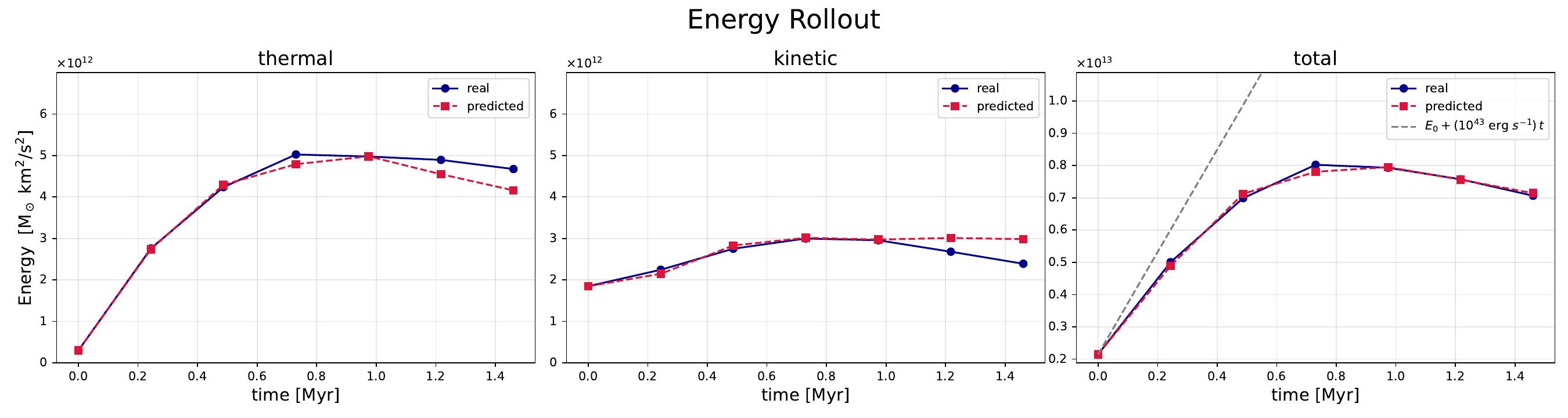}
  \caption{Energy rollouts for the same validation simulation and model realization \xfg\ as Figure \ref{fig:rollouts}. Total energy should never exceed the total initial ISM energy ($E_0$) plus inserted jet-energy $(L_{\text{jet}}t)$, plotted in grey on the total-energy rollout plot (right). For more examples of energy rollouts on an array of validation simulations and different $\Delta t$, see Appendix \ref{sec:appendixrollout}.}
  \label{fig:energyroll}
\end{figure}

\paragraph{Rollout metric.} In Fig. \ref{fig:rollouts}, we plot the rollout for slices of the number density on held-out validation data. Over timesteps of $\Delta t\approx$ 0.25\,Myr, the model predicts 6 snapshots, with each prediction relying on the previous model output for both conditioning and $z_0$. Both the morphology and intensity of $n^*$ are reflected in the intermediate and final snapshots of the model's prediction. The corresponding total energy values for each prediction in the rollout sequence are plotted in Figure \ref{fig:energyroll}, further reflecting the stability of model rollout. For more rollout tests, see Appendix \ref{sec:appendixrollout}.

\paragraph{Limitations and future work.} Our model requires fine-tuning in the low-density, high-jet-power regime, where there are large changes in the physical quantity distributions over short timesteps. In these cases, rollout \err\ and $\varepsilon_{\text{err}}$ can reach several tens of percent, which will require further model optimization to prevent. Importantly, the number of generated snapshots for a given set of simulation initial conditions varies widely; configurations where the jet quickly leaves the $1\,$kpc box contribute much fewer snapshots than slow-moving jets. Thus, to improve generalization, we may also need to generate more training data in this regime.

Once implemented, our model will evolve the central (1\,kpc)$^3$ volume around BHs in cosmological simulations while also supplying the information needed to couple to larger-scale subgrid prescriptions. For example, the ArkenstoneBH model \cite{arkenstone}, which resolves jet interactions with the circumgalactic medium (CGM), couples jet energy at the ISM/CGM boundary. Our model can therefore inform the initial conditions of the ArkenstoneBH model, such as the outflow opening angle, velocity and energy flux at the ISM/CGM boundary. Further, it can also provide a better estimate of the accretion rate onto the BH that accounts for the jet--ISM interaction. Together, this composite sub-grid model will provide physically informed feedback across a wide range of scales in cosmological simulations, from the jet-launching region out to the CGM—ultimately enabling a better understanding of how AGN feedback shapes galaxy evolution.

\bibliographystyle{IEEEtran}
\bibliography{paper}

%%%%%%%%%%%%%%%%%%%%%%%%%%%%%%%%%%%%%%%%%%%%%%%%%%%%%%%%%%%%
\newpage
\appendix
\section*{Appendix}
The structure of the appendix is as follows:
\begin{itemize}
    \item Appendix \ref{sec:appendixa} reviews the data preprocessing methodology, from the original simulations to those used for training and validation.
    \item Appendix \ref{sec:appendixb} gives more information on the training and validation data, including relevant simulation initial conditions.
    \item Appendix \ref{sec:performance} includes a variety of model performance evaluations, including both one step predictions and rollout metrics across different validation simulations.
\end{itemize}

\section{Data Preprocessing}\label{sec:appendixa}

We define aggregate number density and velocity per grid-cell as
\begin{equation}\label{eq:1}
    \Bar{n} = \frac{1}{\Bar{\mu}}\frac{\sum_i \rho_i V_i}{\sum_i V_i}, \quad \Bar{v}_j = \frac{\sum_i m_i v_{ij}}{\sum_i m_i},
\end{equation}
where $\Bar{\mu}=\mu m_p$, $\mu =  0.6165$ is the unitless molecular weight of the simulated gas composition, $m_p$ is the mass of the proton. For the $i$-th particle in a given cell, $\rho_i$ is the mass-density, $V_i$ is the volume, $m_i$ is the mass, and $v_{ij}$ is the velocity component, $v_j\in \{v_x,v_y,v_z\}$, of a particle $i$ in a given cell. In order to provide total energy conservation, we define the internal energy of a grid-cell as
\begin{align}\label{eq:3}
    \Bar{u} = \frac{\sum_i m_i u_i}{\sum_i m_i}+\Delta \text{KE},
\end{align}
where $m_i$ and $u_i$ are the mass and internal energies of a particle $i$ in a given grid-cell. The extra kinetic energy $\Delta$KE term is given by
\begin{align}
    \Delta \text{KE} = \frac{1}{2}\frac{\sum_im_i(v_i^2-\Bar{v}^2)}{\sum_i m_i},
\end{align}
where $\Bar{v}^2 = \Bar{v}_x^2 + \Bar{v}_y^2 +\Bar{v}_z^2$ for one grid-cell. This term accounts for the kinetic energy lost while coarse-graining, as components of velocity cancel while summing in Eq. \ref{eq:1}, and now instead contribute to the grid-cell's internal energy.

Our training data includes a wide range of physical quantity magnitudes representing different energetic jet states. Thus, our final preprocessing step was to normalize each quantity to ensure the model could effectively learn from the dataset, as NNs prefer regularly distributed data. We define normalized number density $n^*$,  internal energy $u^*$, and normalized velocities $v^*_j$ (following the notation of \cite{hirashima}) as 
\begin{align}
    n^* &:= \log_{10}(\Bar{n}),\quad u^* := \frac{1}{N_u}\log_{10}(\Bar{u}) -0.5, \quad v^*_j:=  \frac{\sign(\Bar{v}_j)}{N_v}\log_{10}\left(|\Bar{v}_j|+1\right),
\end{align}
where again, $v_j\in \{v_x,v_y,v_z\}$. Internal energies and velocities have an additional normalization constant $N_u=\log_{10}(8.877\times 10^8)$ and $N_v= \log_{10}(9.2536\times 10^4)$, respectively. These values correspond to the $\log_{10}$ of the maximum velocity and internal energy present in the training set. We also rescale each channel $c$ by a constant factor $a_c \propto 1/\sqrt{E[\Delta X_c^2]}$, where $\Delta X = X_f - X_i$ and the expectation is taken over all training pairs. The factors are normalized such that $\frac{1}{5}\sum_c a_c^2 = 1$, yielding $a = (0.2063,1.9388,0.6300 \times 3)$. This rescaling equalizes the mean squared displacement across channels, so that
each channel contributes equally to the loss. Finally, we subtract 0.5 from the internal energy channel to center the distribution around zero, an adjustment we found empirically improves model performance.  

Due to the nature of Voronoi tessellation, individual Voronoi cells can be large, particularly in diffuse regions. In these diffuse regions, some grid-cells enclose no generating particle. For each such grid-cell, we (1) sample the cell on a uniform 4×4×4 subgrid with 64 sample points, (2) assign each sample point to its generating particle $i$, and (3) estimate the fraction $f_i$ of sample cells assigned to the $i$th particle. We then evaluate Eqs. \ref{eq:1}--\ref{eq:3} over these overlapping cells, instead weighting the number density by $V_if_i$ and the velocities and internal energy by $m_i f_i$. Cells without an interior generating-particle thus inherit the properties of the occupying Voronoi cells.

\section{Training and validation data}\label{sec:appendixb}
\begin{figure}[h!]
  \centering
    \includegraphics[width=0.5\linewidth]{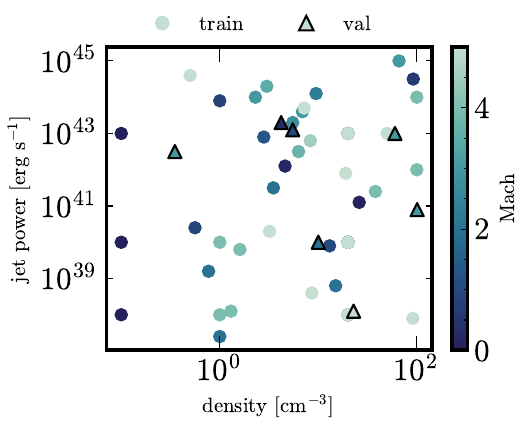}
  \caption{Train vs validation data distribution in the simulation parameter space. The initial conditions include number density and mach number of the ISM, and constant inserted jet power \cite{olga}.}
  \label{fig:split}
\end{figure}

Figure \ref{fig:split} displays the distribution of training and validation simulations across the three initial conditions of number density, mach number, and jet power—in total, we include 7 validation simulations and 50 training simulations in our dataset.

\section{Model performance}\label{sec:performance}

\subsection{One-step predictions}\label{sec:onestep}
To evaluate the model's predictive power over single step predictions, we evaluate our scalar metrics (Equations \ref{eq:deltaerr} \& \ref{eq:energyerr}) across 30 realizations for five validation simulations. Because the model performs differently across small vs. large timesteps, we also include two different $\Delta t$ evaluations for each simulation. The results in Table \ref{tab:valperformance} emphasize the difference in performance across initial conditions. However, as discussed in Section \ref{sec:results}, the metric averaged across all validation pairs yields $\delta_{\text{err}} =  0.94 \pm 1.0 \%$, highlighting the model's general accuracy and generalization abilities. 

\begin{table}[h!]
\centering
\renewcommand{\arraystretch}{1.1}
\caption{Model performance on different validation simulations and timesteps, $\Delta t$, in Myr. The simulation's varied initial conditions include constant jet power  (\jet\ , erg\,s$^{-1}$), and the number density ($n$, cm$^{-3}$) and mach number $\mathcal{M}$ of the surrounding medium. Each configuration's mean and standard deviation is calculated from 30 model realizations of \xfg, with the two outlier validation simulations in the last four rows. The $\delta_{\text{err}, v}$ reported here is averaged across all three velocity channels, in model space.}
\begin{tabular}{lccc|cccc}
\toprule
\jet &$n$& $\mathcal{M}$ & $\Delta t$ & $\delta_{\text{err}, n}$ (\%)  &  $\delta_{\text{err}, u}$ (\%) &  $\delta_{\text{err}, v}$ (\%) & $\varepsilon_{\text{err}}$ ($\%$)  \\ 
\midrule
$10^{40}$& $10$& $2$ & 1 & $0.17 \pm 0.01$  & $ 0.19 \pm0.02$ &$ 0.23 \pm 0.01$&$2.12 \pm 2.18$ \\
$10^{40}$& $10$& $2$ & 2 & $0.15 \pm 0.01$ & $0.15 \pm 0.01$ & $0.22 \pm 0.01$& $2.40 \pm 2.15$\\
\midrule
$10^{43}$& $60$& $3$ & 1 & $3.26 \pm 0.14$ & $6.17 \pm 0.37$ & $4.52 \pm 0.11$ & $-4.08 \pm 4.57$\\
$10^{43}$& $60$& $3$ & 1.5&  $5.63 \pm 0.20$ & $8.64 \pm 0.56$ & $6.58 \pm 0.14$ & $-14.85 \pm 3.97$\\
\midrule
$10^{40.9}$& $100.8$& $3$ & 1& $0.59 \pm 0.01$ & $0.60 \pm 0.02$ & $0.79 \pm 0.01$ & $13.99 \pm 2.83$ \\
$10^{40.9}$& $100.8$& $3$ & 2 & $1.46 \pm 0.03$ & $1.15 \pm 0.02$ & $2.14 \pm 0.03$ & $14.48 \pm 3.21$ \\
\midrule

$10^{38.1}$&  $22.8$& $5.9$ & 1& $1.24 \pm 0.02$& $0.66 \pm 0.01$ & $1.53 \pm 0.02$ &$-31.19 \pm1.64$ \\
$10^{38.1}$&  $22.8$& $5.9$ &2 & $2.14 \pm 0.04$ & $0.87 \pm 0.01$ & $3.14 \pm 0.04$ & $-54.85 \pm 1.46$\\
\midrule
$10^{43.3}$& $4.2$& $0.2$ & 0.25 & $1.66 \pm 0.15$& $37.69 \pm2.58$& $7.77 \pm0.21$ & $-73.10 \pm 4.20$ \\
$10^{43.3}$& $4.2$& $0.2$ & 0.45& $4.37 \pm 0.25$& $34.88 \pm 1.86$&$9.93 \pm0.30$ & $-28.31 \pm 7.29$ \\
\bottomrule\hspace{1mm}
\end{tabular}

\label{tab:valperformance}
\end{table}

\subsection{Rollouts}\label{sec:appendixrollout}
\begin{figure}[h!]
  \centering
    \includegraphics[width=0.97\linewidth]{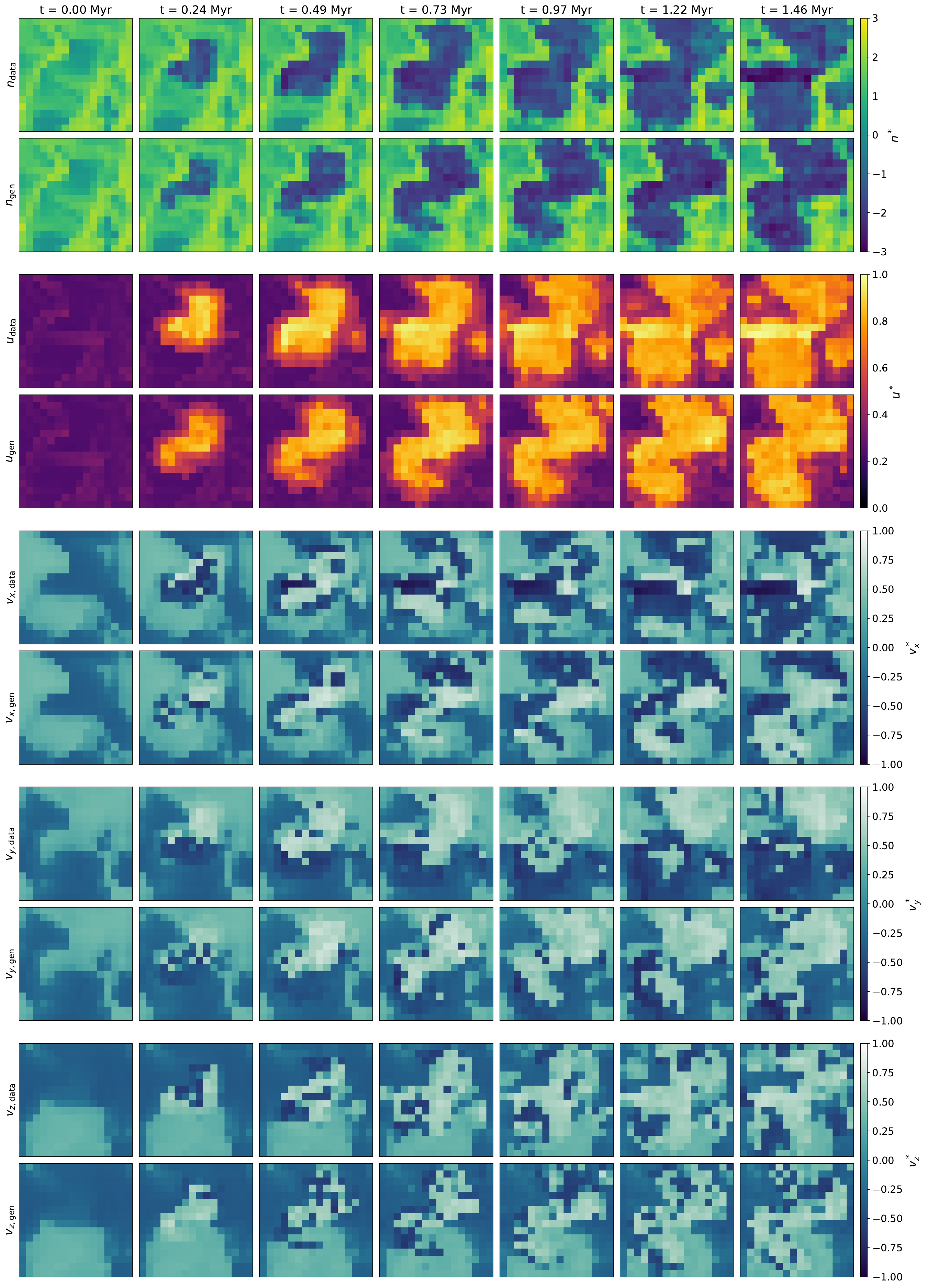}
  \caption{Model rollout for $n=60\,\text{cm}^{-3}, L_{\text{jet}} = 10^{43}\,\text{erg}/\text{s}, \mathcal{M} =3$ validation simulation. Over 6 steps of $\Delta t \approx0.25$\,Myr, the model feeds its own previous output back in as \xi\ to predict the next snapshot, \xfg. From top to bottom, all five physical quantities of number density $n^*$, internal energy $u^*$, and the three velocity components $v_x^*,v_y^*,$ and $v_z^*$ are displayed.}
  \label{fig:allsliceroll}
\end{figure}

Figure \ref{fig:allsliceroll} shows the rollout for the same validation simulation as is displayed in Figure \ref{fig:rollouts} but for all five physical quantities. In all five parameters, we see the general morphologies reconstructed well in each rollout step. However, the worst-performing cells are those with the highest velocities, particularly the negative direction cells in $v_x^*$ and $v_y^*$.

\begin{figure}
  \centering

  \begin{subfigure}[b]{\linewidth}
    \includegraphics[width=0.99\linewidth]{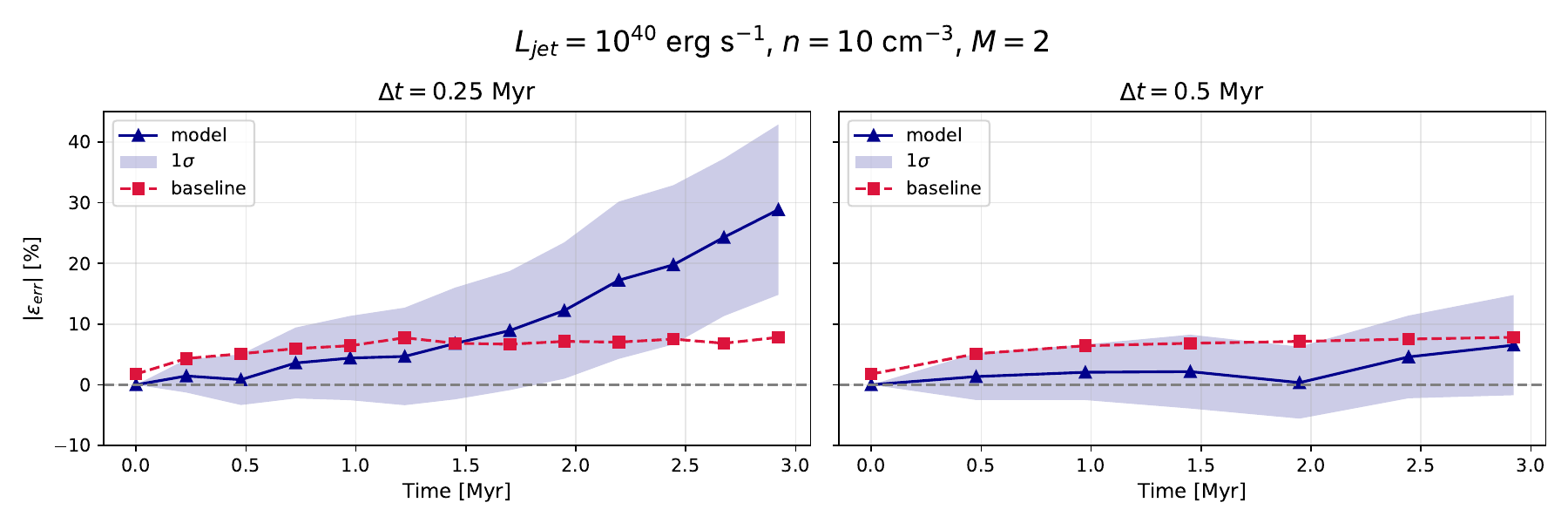}
  \end{subfigure}\hfill
  \begin{subfigure}[b]{\linewidth}
    \includegraphics[width=0.99\linewidth]{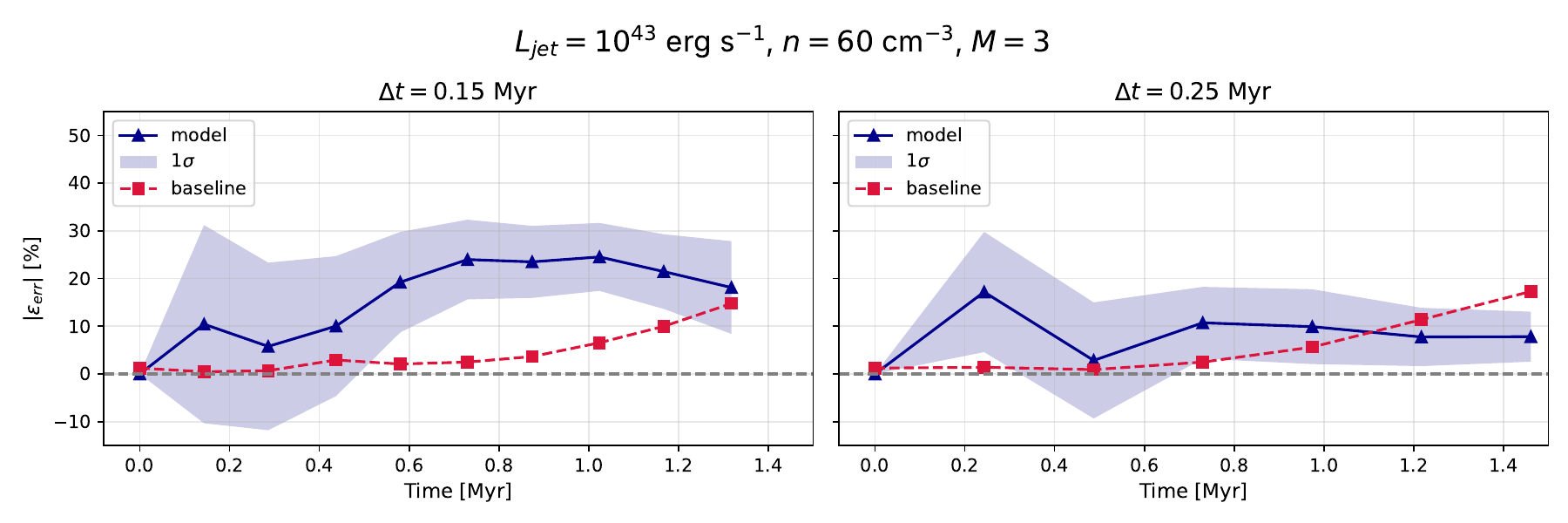}
  \end{subfigure}
    \begin{subfigure}[c]{\linewidth}
    \includegraphics[width=0.99\linewidth]{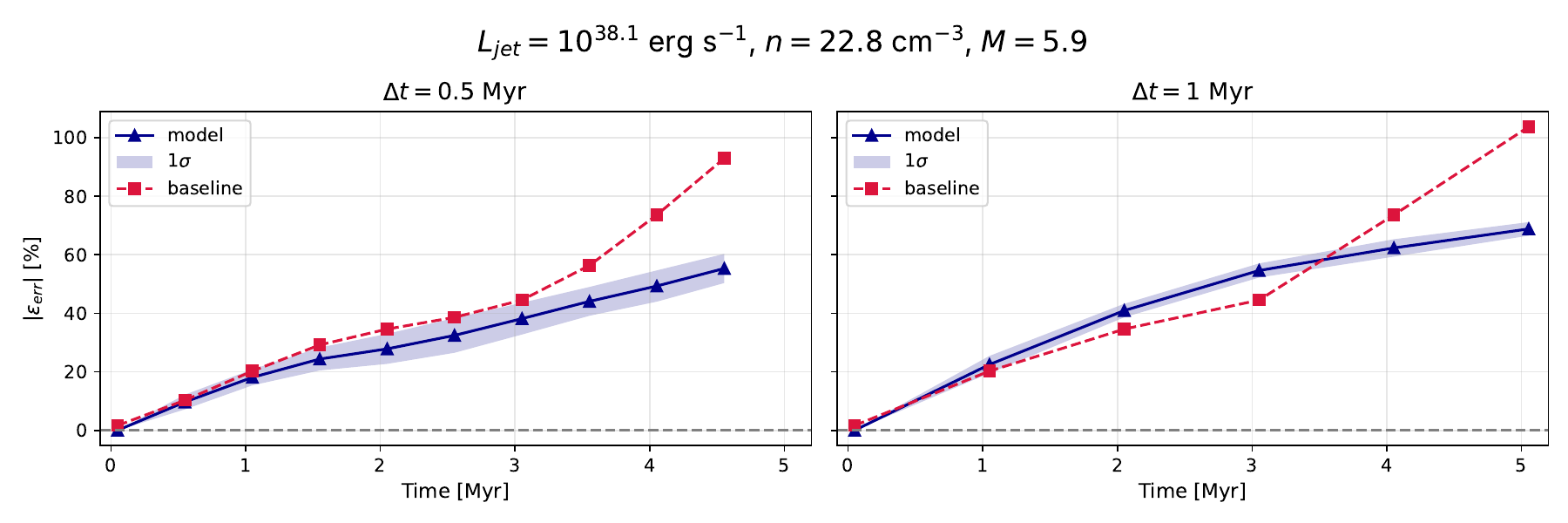}
  \end{subfigure}
    \begin{subfigure}[d]{\linewidth}
    \includegraphics[width=0.99\linewidth]{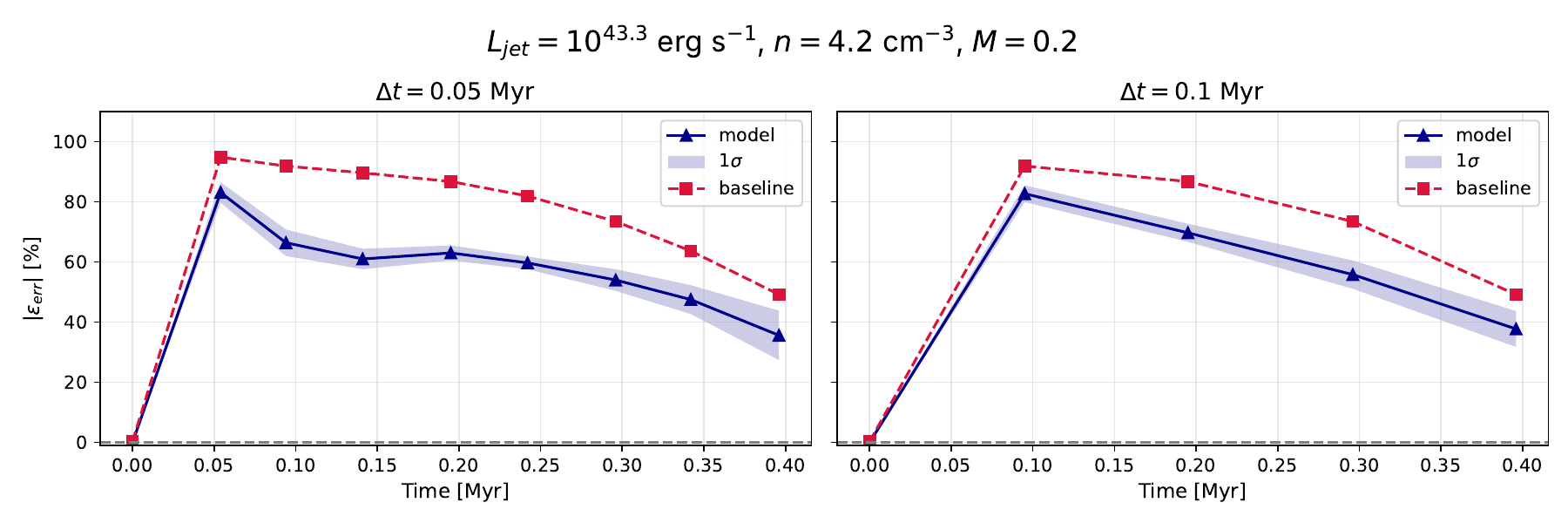}
  \end{subfigure}
    \caption{Total energy rollouts for four validation simulations. The blue line is our model rollout over 20 independent trajectories, with the corresponding mean and standard deviation. The red line is the physical baseline rollout, resulting from direct hydrodynamical simulations on a coarse-grained Voronoi mesh.}
  \label{fig:allenergyrolls}
\end{figure}

For four validation simulations, we calculate the energy performance over rollout with varying timesteps, $\Delta t$, over 20 independent trajectories and average over each. The $\varepsilon_{\text{err}}$ over model rollout for each simulation is displayed in Figure \ref{fig:allenergyrolls}, where the final step's $\varepsilon_{\text{err}}$ is listed in Table \ref{tab:rollout}. There is no universal model preference for many small $\Delta t$ steps or fewer larger $\Delta t$ steps. For the two simulations where the model is performing better (first four rows in Table \ref{tab:rollout}), the model seems to prefer larger steps; For the two simulations where the model is performing worse (last four rows in Table \ref{tab:rollout}), the model seems to prefer smaller steps. In all except two rollouts tested in Table \ref{tab:rollout}, our model performs better than the physical simulation baseline, described below. However, there is still further optimization required, particularly the final two simulations tested in Table \ref{tab:rollout}.

\begin{table}[h!]
\centering
\caption{Model energy rollout performance for four different validation simulations. For each simulation, 20 independent rollout trajectories are evaluated with differing timesteps $\Delta t$ (Myr) and step count. The listed $\varepsilon_{\text{err}}$ is for the final model rollout step, while $\varepsilon_{\text{err}, \text{base}}$ is for the final baseline step, both with respect to the true energy from \xfd. Each listed validation simulation corresponds with a rollout subplot in Figure \ref{fig:allenergyrolls}.}
\begin{tabular}{lcccccc}
\toprule
\jet\ (ergs$\, s^{-1}$) &$n$ (cm$^{-3}$)& $\mathcal{M}$ & $\Delta t$ (Myr) & Step count &   $|\varepsilon_{\text{err}}|$ (\%)& $|\varepsilon_{\text{err}, \text{base}}|$ (\%)\\
\midrule
$10^{40}$ &$ 10$& $2$ &  0.25  & 12 & $28.85\pm 14.03 $ & $7.82$\\
$10^{40}$ &$ 10$& $2$ &   0.5  & 6& $6.53 \pm8.25$ & $7.82$ \\
\midrule
$10^{43}$ &$ 60$& $3$&  0.15  & 9&  $18.11 \pm 9.72$ & $14.77$\\
$10^{43}$ &$ 60$& $3$ &  0.25  & 6& $7.81 \pm 5.24$& $17.30$\\
\midrule
$10^{38.1}$& $22.8$& $ 5.9$ &  0.5 & 9 & $ 55.32 \pm4.98$ & $92.95$\\
$10^{38.1}$& $22.8$& $ 5.9$ &  1 & 5 & $68.83 \pm 2.32$& $103.74$\\
\midrule
$10^{43.3}$& $4.2$&$0.2$ &  0.05 & 8 & $ 35.61 \pm 8.21$ & $49.7$\\
$10^{43.3}$& $4.2$&$0.2$ &  0.1 & 4 & $37.73 \pm 5.98$ & $49.7$\\
\bottomrule\hspace{1mm}
\end{tabular}
\label{tab:rollout}
\end{table}

As a physical baseline, we run lower-resolution jet-ISM hydrodynamical simulations in AREPO. We first take a high-resolution snapshot before the jet turns on, then coarse-grain to a Voronoi mesh with mass resolution of $3\times 10^4 M_\odot$, similar to the baryonic mass resolution of TNG50 \cite{tng50}. The unresolved kinetic energy is added to the thermal energy, similar to the pre-processing steps outlined in Appendix \ref{sec:appendixa}. We then turn on the jet and perform the hydrodynamical simulation; jet refinement is kept so jet resolution does not get lower than (30\,pc)$^3$. The new, lower-resolution baseline simulations are then projected onto a cartesian grid in the same manner as the original, high-resolution training simulations, as detailed in Appendix \ref{sec:appendixa}. 

When Voronoi coarse-graining the baseline simulations, the added thermal energy is quickly radiated away before the initial snapshot is recorded. Thus, when assessing the baselines and plotting in Figure \ref{fig:allenergyrolls}, we add the $\Delta E_{\text{kin}}$ between the initial baseline and original snapshots as a constant factor to all snapshots. All reported errors and plots with baseline energies include this factor; in the future, we will add a sub-grid model for the unresolved turbulence similar to \cite{semenov2025}, so that this energy doesn't radiate away so quickly.

\end{document}